\documentclass[%
 reprint,
superscriptaddress,
 amsmath,amssymb,
 aip,
floatfix,
]{revtex4-2}

\usepackage{graphicx}
\usepackage{dcolumn}
\usepackage{bm}
\usepackage{times,here}
\usepackage{braket}
\usepackage{ascmac}
\usepackage{amsmath,amssymb,empheq}
\usepackage{nameref} 
\usepackage{hyperref}
\usepackage{color}
\newcommand{\e}{\textrm{e}}
\newcommand{\im}{\textrm{i}}
\usepackage[usenames,dvipsnames]{xcolor}

\begin{document}
 
\preprint{APS/123-QED}

\title{High-performance conditional-driving gate for Kerr parametric oscillator qubits}

\author{Hiroomi Chono}
\email{hiroomi1.chono@toshiba.co.jp}
\affiliation{Frontier Research Laboratory, Corporate Research \& Development Center, Toshiba Corporation, Saiwai-ku, Kawasaki 212-8582, Japan}
\author{Hayato Goto}
\affiliation{Frontier Research Laboratory, Corporate Research \& Development Center, Toshiba Corporation, Saiwai-ku, Kawasaki 212-8582, Japan}
\affiliation{RIKEN Center for Quantum Computing (RQC), Wako, Saitama 351-0198, Japan}

\date{\today}

\begin{abstract}
Kerr parametric oscillators (KPOs), two-photon driven Kerr-nonlinear resonators, can stably hold coherent states with opposite-sign amplitudes and are promising devices for quantum computing. 
Recently, we have theoretically proposed a two-qubit gate $R_{zz}$ for highly detuned KPOs and called it a conditional-driving gate [Chono \textit{et al}., Phys. Rev. Res. \textbf{4}, 043054 (2022)]. 
In this study, analyzing its superconducting-circuit model and deriving a corresponding static model, we find that an AC-Zeeman shift due to the flux pulse for the gate operation largely affects the gate performance.
This effect becomes a more aggravating factor with shorter gate times, leading to an increase in the error rate. 
We thus propose a method to cancel this undesirable effect. 
Furthermore, through the use of shortcuts to adiabaticity and the optimization of flux pulses, we numerically demonstrate a conditional-driving gate with average fidelity exceeding 99.9\% twice faster than that without the proposed cancellation method and the STA.
\end{abstract}

\maketitle

\section{Introduction}\label{Introduction}
Kerr parametric oscillators (KPOs) can stabilize quantum superpositons of two coherent states with opposite-sign amplitudes. 
The quantum-mechanical superposition states are also known as Schr\"{o}dinger cat states. 
In recent years, quantum computations using the coherent states as computational basis states, known as KPO qubits or Kerr-cat qubits, have been intensively researched~\cite{Goto2016,Goto2016a,Puri2017a,Goto2019a}.  
The KPO qubits have garnered attention in recent years because of their applicability to both 
quantum annealing~\cite{Goto2016,Goto2019a,Nigg2017,Puri2017,Goto2018a,Onodera2020,Goto2020b,Kanao2021} and 
gate-based quantum computing~\cite{Goto2016a,Goto2019a,Puri2017a,Darmawan2021a,Kwon2022,Masuda2022a,Kanao2022b,Xu2022,Chono2022,Aoki2023,Kang2022,Kang2023,Kanao2024}.
Since coherent states are robust against photon loss, the KPO qubits can suppress bit-flip errors~\cite{Puri2017a,Cochrane1999}. 
The KPO qubits can be implemented using superconducting circuits~\cite{Wang2019,Grimm2020,Kwon2022,Iyama2024,Hoshi2024}, and the suppression of bit-flip errors has been demonstrated experimentally~\cite{Grimm2020}. 

A relatively easy-to-implement universal quantum gate set for KPO qubits consists of single-qubit gates, $R_x$ and $R_z$, and a two-qubit gate $R_{zz}$~\cite{Goto2016a}. 
Adiabatic single-qubit $R_z$ gates and a nonadiabatic $R_x$ gate have been realized experimentally~\cite{Grimm2020}. 
In addition, various gate implementations for KPO qubits have been theoretically proposed~\cite{Masuda2022a,Kanao2022b,Chono2022,Aoki2023,Kang2022,Kang2023}. 
In particular, the acceleration of gate operations has been actively studied~\cite{Xu2022,Kanao2024}. 
In the case of KPO qubits, single-photon loss is a dominant source of errors, necessitating short gate times, but then nonadiabatic processes can cause transitions out of the qubit subspace. 
To reduce the errors, the accelerations of gates by numerically optimizing flux-pulse waveforms and by applying shortcuts to adiabaticity (STA) using counterdiabatic terms~\cite{Kanao2024,Guery-Odelin2019} have been proposed~\cite{Xu2022,Kanao2024}. 
However, the previous studies are based on models simplified through several approximations, rather than rigorous superconducting-circuit models.

On the other hand, we theoretically proposed an $R_{zz}$ gate for KPO qubits utilizing the three-wave mixing process induced by the sum-frequency microwave drive, which we named a conditional-driving (CD) gate~\cite{Chono2022}.
This gate can be implemented using a simple superconducting circuit where two KPO qubits with large detuning are coupled via a capacitor, without employing a tunable coupler. 
Numerical simulations of the gate operation using its superconducting-circuit model demonstrated that $R_{zz}$ gates with fidelity over 99.9\% can be achieved in the rotation angle region necessary for universal quantum computation. 
In the theoretical proposal, however, the high performance required a long gate time for adiabatic operations, and the flux-pulse waveform was not optimized. 

In this study, we attempt to achieve faster gate operations while keeping high fidelity by adding a counterdiabatic term for the STA and optimizing all the flux-pulse waveforms.
We first analyze the superconducting-circuit model and derive a rotating-wave approximation (RWA) model, leading to AC-Zeeman shifts induced by parametric pumps, a flux pulse for the gate operation, and the counterdiabatic term. 
We compensate the undesired shifts using additional fluxes. 
Next, we derive a static model and consequently find an additional AC-Zeeman shift. 
We thus propose a method to cancel the second undesired shift by an additional flux pulse, which we call a cancellation term. 
We numerically show that the cancellation term together with the flux pulse optimization and the STA allows us to achieve
our goal, i.e., the above-mentioned acceleration, which cannot be achieved without the cancellation term.

This paper is organized as follows. 
In Sec.~II, we introduce the CD gate and its superconducting-circuit model, derive a RWA model, and present the CD-gate optimization without the cancellation term. 
In Sec.~III, we derive a static model resulting in the second AC-Zeeman shift, and show the dramatic improvement of CD-gate performance with the cancellation term.
Finally, we conclude this study in Sec.~IV.

\section{CD gate and models}\label{opt} 
\begin{figure}
\includegraphics[width=8.5cm]{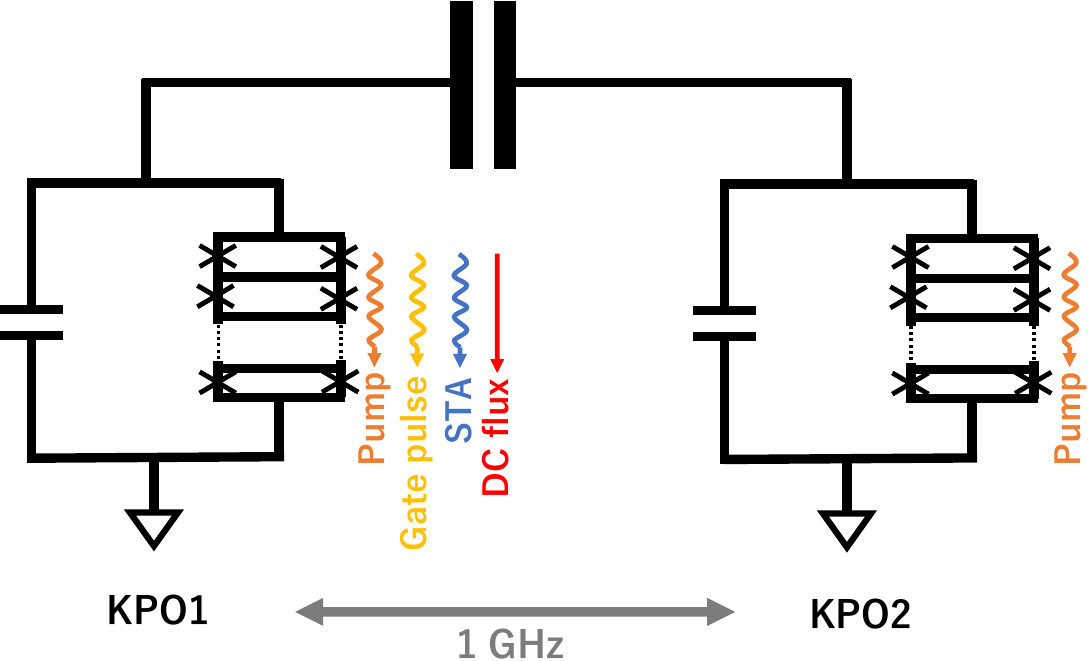}
\caption{\label{fig:circuit}
Circuit diagram of two KPOs for the $R_{zz}$ gate and its acceleration. 
Two KPOs are implemented as transmons with a DC-SQUID array~\cite{Wang2019}.
The detuning frequency is set to 1~GHz.
} 
\end{figure}

\subsection{Superconducting-circuit model}
We first explain the CD gate for KPO qubits implemented with a superconducting circuit shown in Fig.\ref{fig:circuit}, where two highly detuned KPOs consist of a shunt capacitor and an array of DC superconducting quantum interference devices (SQUIDs), and are coupled via a capacitor. 
The parametric pumps for the KPOs are realized by AC fluxes applied to their DC SQUIDs, as shown in Fig.~{\ref{fig:circuit}}, where each pump frequency is set around twice the eigenfrequency of the corresponding KPO.
The CD gate can be performed by applying an additional AC flux to KPO1, where the drive frequency is set to the sum of the eigenfrequencies of the two KPOs.
This drive generates photons with the phases corresponding to coherent states in KPO1.
The photons are transferred to KPO2 through the capacitor, which thereby enable the manipulation of KPO2 depending on the states in KPO1, resulting in the $R_{zz}$ gate\cite{Chono2022}.

To accelerate the CD gate, we introduce a counterdiabatic term for the STA and optimize the waveforms of the flux pulses to maximize the average gate fidelity using a superconducting-circuit model.
Then, the superconducting-circuit model of the CD gate is given by 
\begin{align}
H \label{eq:SC_model}
&=\sum_{j=1,2}H_j+V,  \\
H_j \label{eq:SC_model2}
&=\omega_j{a_j}^{\dagger}a_j
-\frac{\tilde{E}_{\textrm{J}j}}{2N}\varphi_j^2  \nonumber \\
&\quad
-NE_{\textrm{J}j}\cos\left[\theta_0-\delta_{\textrm{m}j}(t)\right]\cos\frac{\varphi_j}{N}, \\
V \label{eq:SC_model3}
&=\frac{8E_{\textrm{C}1}E_{\textrm{C}2}}{E_{\textrm{C}}+E_{\textrm{C}1}+E_{\textrm{C}2}}n_1n_2, 
\end{align}
where $H_j$ is the Hamiltonian of KPO$j~(j=1,2)$,
$V$ is the interaction Hamiltonian between the two KPOs,
$E_{\textrm{C}j}(E_{\textrm{C}})$ is the charging energy of the shunt capacitor (coupling capacitor), 
$E_{\textrm{J}j}$ is the Josephson energy, 
$a_j$ is an annihilation operator,
$n_{j}$ and $\varphi_{j}$ are the Cooper-pair number and phase-difference operators, respectively,
$\theta_0$ and $\delta_{\textrm{m}j}$ are the angles corresponding to DC and modulated fluxes, respectively,
$N$ is the number of the DC SQUIDs in the array, 
and $\tilde{E}_{\textrm{J}j}(=E_{\textrm{J}j}\cos\theta_0)$ is the effective Josephson energy.
We also set the reduced Planck constant $\hbar$ to 1, and define
$\omega_j, n_j$, and $\varphi_j$, satisfying the commutation relation $[\varphi_j, n_j]=\im$, as
\begin{align}
\omega_j& \label{eq:omega}
=\left(\frac{8E_{\textrm{C}j}\tilde{E}_{\textrm{J}j}}{N}\right)^{\frac{1}{2}}, \\
n_j&={\im}\left(\frac{\tilde{E}_{\textrm{J}j}}{32NE_{\textrm{C}j}}\right)^\frac{1}{4}
(a_j^\dagger-a_j), \\
\varphi_j&=\left(\frac{2NE_{\textrm{C}j}}{\tilde{E}_{\textrm{J}j}}\right)^\frac{1}{4}(a_j^\dagger+a_j).
\end{align}    

We set the angles for the modulated fluxes in Eq.~(\ref{eq:SC_model2}) as 
\begin{align}  
\delta_\textrm{m1}(t) 
&=\delta_1\cos(\omega_{\textrm{p}1}t) \nonumber \\
&\quad
+\delta_\textrm{g}(t)\cos(\omega_\textrm{g}t)+{\delta}_\textrm{g}^{\prime}(t)\sin(\omega_\textrm{g}t), \label{eq:d1} \\
\delta_\textrm{m2}(t)
&=\delta_2\cos(\omega_{\textrm{p}2}t), \label{eq:d2}
\end{align} 
where $\delta_j~(j=1,2), \delta_\textrm{g},$ and $\delta_\textrm{g}^{\prime}$ are the amplitudes corresponding to the parametric pump for KPO$j$, the gate pulse, and the counterdiabatic term~\cite{Kanao2024}, respectively, 
and $\omega_{\textrm{p}j}$ and $\omega_\textrm{g}$ are the frequencies of the parametric pump for KPO$j$ and the gate pulse, respectively.

In this work, we set the parameters as ${\omega_1/(2\pi)=10}$~GHz, ${\omega_2/(2\pi)=11~\textrm{GHz}}$, ${\theta_0=\pi/4}$, ${N=5}$, and ${E_{\textrm{C}j}/(2\pi)=300}$~MHz.
Using these values, we can determine $E_{\textrm{J}j}$ by Eqs.~(\ref{eq:omega}) and also ${\tilde{E}_{\textrm{J}j}=E_{\textrm{J}j}\cos\theta_0}$.

\subsection{Rotating-wave approximation model}
To obtain the relations between KPO and superconducting-circuit parameters, here we derive an RWA model from the superconducting-circuit model in Eqs.~(\ref{eq:SC_model}-\ref{eq:SC_model3}) based on several approximations, which is useful for parameter settings.
First, the Hamiltonian $H_1$ can be rewritten and approximated as
\begin{widetext}
\begin{equation}\begin{split}\label{eq:RWA_model2}
H_1
&=\omega_1a_1^{\dagger}a_1
-\frac{\tilde{E}_{\textrm{J}1}}{2N}\varphi_1^2 
-NE_{\textrm{J}1}\left[\cos\theta_0\cos\delta_\textrm{m1}
+\sin\theta_0\sin\delta_\textrm{m1}\right]\cos\frac{\varphi_1}{N} \\
&\rightarrow
\omega_1a_1^{\dagger}a_1
-\frac{\tilde{E}_{\textrm{J}1}}{2N}\varphi_1^2 
-N\tilde{E}_{\textrm{J}1}\left(1-\frac{\delta_\textrm{m1}^2}{2}\right)
\left(-\frac{\varphi_1^2}{2N^2}+\frac{\varphi_1^4}{24N^4}\right)
+N\tilde{E}_{\textrm{J}1}\tan\theta_0\cdot\delta_\textrm{m1}\frac{\varphi_1^2}{2N^2},
\end{split}\end{equation}\label{eq:RWA_model3}
\end{widetext}
where we have taken the transmon limit as 
${\cos(\varphi_1/N)\rightarrow-\varphi_1^2/(2N^2)+\varphi_1^4/(24N^3)}$, have used the approximations
${\cos\delta_\textrm{m1}\simeq 1-\delta_{m1}^2/2}$ and ${\sin\delta_\textrm{m1}\simeq\delta_\textrm{m1}}$, and have dropped the tiny terms including $\delta_{\textrm{m}1}\varphi_1^4$. 

Second, moving into the rotating frame at the frequency $\tilde{\omega}_1$, where $\tilde{\omega}_j$ is the eigenfrequency of KPO$j~(j=1,2)$ obtained by numerically diagonalizing $H$ with ${\delta_{\textrm{m}j}=0}$ in Eqs.~(\ref{eq:SC_model}-\ref{eq:SC_model3}), we obtain
\begin{widetext}
\begin{equation}\begin{split}\label{eq:RWA_model3}
H_1
&=
(\omega_1-\tilde{\omega}_1)a_1^{\dagger}a_1
-\frac{E_{\textrm{C}1}}{12N^2}(a_1^\dagger\e^{\im\tilde{\omega}_1t}+a_1\e^{-\im\tilde{\omega_1}t})^4
+\sqrt{\frac{\tilde{E}_{\textrm{J}1}E_{\textrm{C}1}}{2N}}\tan\theta_0\delta_\textrm{m1}
(a_1^\dagger\e^{\im\tilde{\omega}_1t}+a_1\e^{-\im\tilde{\omega}_1t})^2 \\
&\quad
-\frac{1}{2}\sqrt{\frac{\tilde{E}_{\textrm{J}1}E_{\textrm{C}1}}{2N}}\delta_\textrm{m1}^2
(a_1^\dagger\e^{\im\tilde{\omega}_1t}+a_1\e^{-\im\tilde{\omega}_1t})^2.
\end{split}\end{equation}
\end{widetext}

Third, substituting Eq~(\ref{eq:d1}) with $\omega_{\textrm{p}j}=2\tilde{\omega}_j$ and
$\omega_\textrm{g}=(\omega_{\textrm{p}1}+\omega_{\textrm{p}2})/2$
and performing the RWA in Eq.~(\ref{eq:RWA_model3}), namely, neglecting the oscillating terms faster than the detuning ${\Delta}_{12}(\equiv\tilde{\omega}_1-\tilde{\omega}_2)$, 
we obtain
\begin{widetext}
\begin{align}\label{eq:H1}
H_1
&\simeq
(\omega_1-\tilde{\omega}_1)a_1^{\dagger}a_1
-\frac{E_{\textrm{C}1}}{N^2}a_1^{\dagger}a_1
-\frac{1}{2}\sqrt{\frac{\tilde{E}_\textrm{J1}E_{\textrm{C}1}}{2N}}
(\delta_1^2+\delta_\textrm{g}^2+\delta_\textrm{g}^{\prime2})a_1^{\dagger}a_1
-\frac{E_{\textrm{C}1}}{2N^2}a_1^{\dagger2}a_1^2 
\nonumber \\
&\quad
+\sqrt{\frac{\tilde{E}_{\textrm{J}1}E_{\textrm{C}1}}{2N}}\tan\theta_0
\left[
\frac{\delta_1}{2}(a_1^{\dagger2}+a_1^2)
+\frac{\delta_\textrm{g}}{2}(a_1^{\dagger2}\e^{\im\Delta_\textrm{12}t}+a_1^2\e^{-\im\Delta_\textrm{12}t})
+\frac{\delta_\textrm{g}^\prime}{2\im}(a_1^{\dagger2}\e^{\im\Delta_\textrm{12}t}-a_1^2\e^{-\im\Delta_\textrm{12}t})
\right] \nonumber \\
&\equiv
\Delta_1a_1^{\dagger}a_1
-\frac{K_1}{2}a_1^{\dagger2}a_1^2
+\frac{P_1}{2}(a_1^{\dagger2}+a_1^2)
+\frac{p_\textrm{g}}{2}(a_1^{\dagger2}\e^{\im\Delta_\textrm{12}t}+a_1^2\e^{-\im\Delta_\textrm{12}t})
+\frac{p_\textrm{g}^\prime}{2\im}(a_1^{\dagger2}\e^{\im\Delta_\textrm{12}t}-a_1^2\e^{-\im\Delta_\textrm{12}t}),
\end{align}
\end{widetext}
where 
\begin{align}    
\Delta_1
&=
\omega_1-\tilde{\omega}_1
-\frac{E_{\textrm{C}1}}{N^2}
-\frac{1}{2}\sqrt{\frac{\tilde{E}_\textrm{J1}E_{\textrm{C}1}}{2N}}
(\delta_1^2+\delta_\textrm{g}^2+\delta_\textrm{g}^{\prime2}), \label{eq:KPO1}\\
K_1  \label{eq:K1}
&=\frac{E_{\textrm{C}1}}{N^2}, \\
P_1 \label{eq:P1}
&=\delta_1\sqrt{\frac{\tilde{E}_{\textrm{J}1}E_{\textrm{C}1}}{2N}}\tan\theta_0, \\
p_\textrm{g} \label{eq:pg}
&=\delta_\textrm{g}\sqrt{\frac{\tilde{E}_{\textrm{J}1}E_{\textrm{C}1}}{2N}}\tan\theta_0,~
p_\textrm{g}^\prime
=\delta_\textrm{g}^\prime\sqrt{\frac{\tilde{E}_{\textrm{J}1}E_{\textrm{C}1}}{2N}}\tan\theta_0. 
\end{align}

Similarly, we obtain
\begin{align}\label{eq:H2}
H_2
&\simeq
\Delta_2a_2^{\dagger}a_2
-\frac{K_2}{2}a_2^{\dagger2}a_2^2
+\frac{P_2}{2}(a_2^{\dagger2}+a_2^2), \\
V \label{eq:V}
&\simeq
g(a_1^{\dagger}a_2\e^{\im\Delta_{12}t}+a_1a_2^{\dagger}\e^{-\im\Delta_{12}t}),
\end{align}
where
\begin{align}
\Delta_2
&=
\omega_2-\tilde{\omega}_2
-\frac{E_{\textrm{C}2}}{N^2}
-\frac{1}{2}\sqrt{\frac{\tilde{E}_\textrm{J2}E_{\textrm{C}2}}{2N}}
\delta_2^2, \label{eq:KPO2} \\
K_2 \label{eq:K2}
&=\frac{E_{\textrm{C}2}}{N^2}, \\
P_2 \label{eq:P2}
&=\delta_2\sqrt{\frac{\tilde{E}_{\textrm{J}2}E_{\textrm{C}2}}{2N}}\tan\theta_0, \\
g\label{eq:g}
&=\frac{2E_\textrm{C1}E_\textrm{C2}}{E_\textrm{C}+E_\textrm{C1}+E_\textrm{C2}}
\left(\frac{\tilde{E}_{\textrm{J}1}\tilde{E}_{\textrm{J}2}}{4N^2E_{\textrm{C}1}E_{\textrm{C}2}}\right)^{\frac{1}{4}}.
\end{align}
We refer to Eqs.~(\ref{eq:H1}-\ref{eq:g}) as the RWA model.
From Eqs.~(\ref{eq:K1}) and (\ref{eq:K2}), we can determine the Kerr coefficients as $K_j/(2\pi)=12$~MHz. 
Also, we set the parametric pump amplitudes as $P_j=4K_j$ so that average photon numbers approximately become 4 during idle time, which determines $\delta_j$ through Eqs.~(\ref{eq:P1}) and (\ref{eq:P2}). 
We also set the couping strength as $g/(2\pi)=10$~MHz, which determines $E_\textrm{C}$ through Eq.~(\ref{eq:g}).

We apply a time-dependent flux to each loop in the DC-SQUID array to cancel the undesired AC-Zeeman shifts in Eqs.~(\ref{eq:KPO1}) and (\ref{eq:KPO2}).
This implementation corresponds to introducing additional angles $\theta_j$ to Eq.~(\ref{eq:SC_model2}) as
\begin{align}\label{eq:cancellation}    
\theta_0-\delta_{\textrm{m}j}(t)&\rightarrow\theta_0-\delta_{\textrm{m}j}(t)-\theta_j(t), \\
\theta_1(t)&
=\frac{\delta_1^2+\delta_\textrm{g}(t)^2+{\delta}_\textrm{g}^{\prime}(t)^2}{4\tan\theta_0}, \nonumber \\
\theta_2(t)&
=\frac{\delta_2^2}{4\tan\theta_0}. \nonumber
\end{align}

We have numerically found that tiny detunings remain even though Eq.~(\ref{eq:cancellation}) is applied.
To compensate the tiny detunings, we adjust the pump frequencies as ${\omega_{\textrm{p}j}=2\tilde{\omega}_j+\Delta_{\textrm{p}j}}$, where $\Delta_{\textrm{p}j}$ are set as ${\Delta_{\textrm{p}1}/(2\pi)=1.9}$~MHz and ${\Delta_{\textrm{p}2}/(2\pi)=1.7}$~MHz. 
\subsection{Gate simulation}
Using the Hamiltonian in Eqs.~(\ref{eq:SC_model}-\ref{eq:SC_model3}) and QuTiP\cite{QuTiP,QuTiP2}, we solve the Schr\"{o}dinger equation with the four initial states set to the computational basis states $\ket{\bar{0}\bar{0}}, \ket{\bar{0}\bar{1}}, \ket{\bar{1}\bar{0}}$ and $\ket{\bar{1}\bar{1}}$ (see Appendix~\ref{sec:appendA} for the definitions of the computational basis states).
We then calculate the average gate fidelity at each gate time  as~\cite{Pedersen2007}
\begin{equation}\label{eq:Fbar}
\begin{split}    
\bar{F}&=\frac{|\textrm{tr}[R_{zz}^{\dagger}(\frac{\pi}{2})U]|^2+\textrm{tr}(UU^\dagger)}{20}, \\
R_{zz}\left(\frac{\pi}{2}\right)
&=
\begin{pmatrix}
1 & 0 & 0 & 0 \\
0 & \e^{\im\frac{\pi}{2}} & 0 & 0 \\
0 & 0 & \e^{\im\frac{\pi}{2}} & 0 \\
0 & 0 & 0 & 1
\end{pmatrix},
\end{split}
\end{equation} 
where $R_{zz}(\frac{\pi}{2})$ represents the ideal $R_{zz}$ gate operation with a rotation angle of $\pi/2$.
The ${4\times4}$ matrix $U$ is defined as follows $(i,j,i^{\prime},j^{\prime}\in\{0,1\})$:
\begin{align}\label{eq:U}
U_{2i+j, 2i^{\prime}+j^{\prime}}
=\braket{\bar{i},\bar{j}|\widetilde{\bar{i}^{\prime},\bar{j}^{\prime}}},
\end{align}    
where $\ket{\widetilde{\bar{i},\bar{j}}}$ is the resultant state of the gate operation on $\ket{\bar{i},\bar{j}}$.

For optimization we express the waveforms of the original gate pulse and the counterdiabatic term as~\cite{Martinis2014,Kanao2024}
\begin{align}
\delta_\textrm{g}(t)
&=\sum_{n=1}^{N_\textrm{f}}\frac{A_n}{2}\left(1-\cos\frac{2n\pi t}{T_\textrm{g}}\right), \\
{\delta}_\textrm{g}^{\prime}(t)
&=\sum_{n=1}^{N_\textrm{f}}B_nn\sin\frac{2n\pi t}{T_\textrm{g}}, 
\end{align}     
where $A_n$ and $B_n$ are the parameters characterizing the waveforms,
$T_\textrm{g}$ is a gate time, and $N_\textrm{f}$ determines the number of frequency components.
The counterdiabatic term is related to a derivative removal by adiabatic gate (DRAG) technique~\cite{Xu2022}.
In Ref.~\cite{Xu2022}, the shapes of the flux pulses are determined analytically, whereas in our calculations, they are optimized numerically.
In this study, we set $N_\textrm{f}$ to only 2.  
This setting is expected to be experimentally feasible.

\begin{figure}
\includegraphics[width=0.45\textwidth]{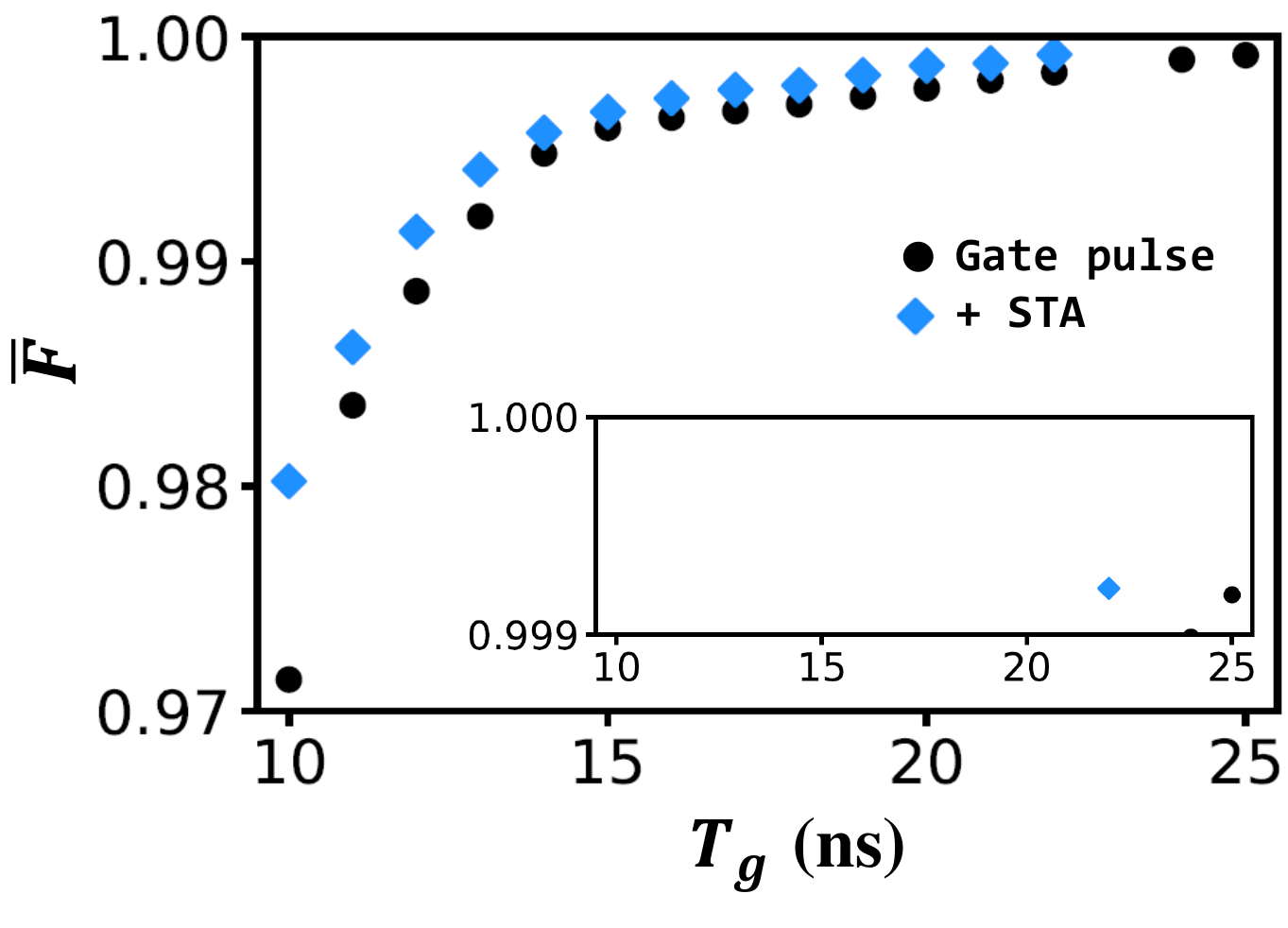}
\caption{\label{fig:AB}
Average gate fidelity of the CD gate with optimized flux pulses at each gate time.
Black circles represent the results with the original gate pulse alone and blue diamonds do the results with a counterdiabatic term for the STA.  
The inset shows a magnified view of $\bar{F}>$99.9\%.
} 
\end{figure}

We numerically optimize the parameters $A_n$ and $B_n$ of the pulse waveforms to maximize the average gate fidelity $\bar{F}$ in Eq.~(\ref{eq:Fbar}), using the optimizer based on the quasi-Newton method with the $L$-BFGS-$B$ formula in the optimparallel package~\cite{Gerber,Gerber2}.
We set the maximum photon number to 20.
We perform the optimization for different gate times from 10~ns to 25~ns, and the results are shown in Fig.~\ref{fig:AB}. 
Figure~\ref{fig:AB} shows the average gate fidelities for the original gate pulse only (black circles) and for that with the counterdiabatic term for the STA (blue diamonds). 
Overall, the result shows minor improvements even with the counterdiabatic term for the STA.
Focusing on the average gate fidelities exceeding 99.9\%, the optimization without the counterdiabatic term achieves a gate time of 25 ns, whereas the use of the counterdiabatic term results in a gate time of 22 ns (see the inset of Fig.~\ref{fig:AB}). 
That is, the STA offers only 10\% acceleration.
\section{Gate optimization with proposed method}\label{opt2}
 
\subsection{Static model}
To investigate the reason why the STA works weakly, we derive a static model from the RWA model [Eqs.~(\ref{eq:H1}-\ref{eq:g})].
The RWA model can be rewritten as
\begin{equation}\label{eq:RWA_model}  
\begin{split}
H_\textrm{RWA}
&=H_\textrm{KPO}+O_\textrm{t}\e^{-\im\Delta_{12}t}+O_\textrm{t}^{\dagger}\e^{+\im\Delta_{12}t}, \\
H_\textrm{KPO}&=\sum_{j=1,2}\left[-\frac{K_j}{2}a_j^{\dagger 2}a_j^2+\frac{P_j}{2}(a_j^{\dagger 2}+a_j^2)\right], \\
O_\textrm{t}&=ga_1a_2^\dagger+\frac{p_\textrm{g}(t)+p_\textrm{g}^\prime(t)}{2}a_1^2.
\end{split}
\end{equation}
Assuming that the gate time is sufficiently longer than $\Delta_{12}^{-1}$, 
then the the static model (see Appendix~\ref{sec:appendB} for the derivation) is given by
\begin{equation}\begin{split}\label{eq:stat_model}
H_\textrm{stat}
&=H_\textrm{KPO}+\frac{\left[O_\textrm{t}, O_\textrm{t}^{\dagger}\right]}{\Delta_{12}}+\mathcal{O}(\Delta_{12}^{-2}), \\
\left[O_\textrm{t}, O_\textrm{t}^{\dagger}\right]
&=g^2(a_1^{\dagger}a_1-a_2^{\dagger}a_2) \\
&\quad-gp_\textrm{g}(t)(a_1^{\dagger}a_2^{\dagger}+a_1a_2)+\im gp_\textrm{g}^{\prime}(t)(a_1^{\dagger}a_2^{\dagger}-a_1a_2) \\
&\quad-\left[p_\textrm{g}^2(t)+p_\textrm{g}^{\prime2}(t)\right](a_1^{\dagger}a_1+\tfrac{1}{2}).
\end{split}
\end{equation} 
In this model, we find that the fourth term in the commutator represents a time-depending detuning, which can be interpreted as an additional AC-Zeeman shift depending on the strength of the gate pulse and the counterdiabatic term. 
Shorter gate times necessitate larger $p_\textrm{g}$ and $p_\textrm{g}^\prime$, which in turn enhances the undesired shift. 
This additional shift may be the reason why the STA works only weakly in Sec.~IIC.
We expect that by canceling this undesired shift, the gate error rate can be reduced. 
Acutually, the acceleration of the $R_{zz}$ gate using Eq.~(\ref{eq:stat_model}) dropping the time-depending detuning term has been numerically achieved~\cite{Kanao2024}.

We also find that the time-periodic beam-splitter-type interaction in Eq.~(\ref{eq:V}) effectively works as a static two-mode squeezing-type interaction. 
This is natural because the CD-gate operation is performed by the sum-frequency microwave drive. 
On the other hand, an $R_{zz}$ gate using difference-frequency one has also been proposed recently~\cite{Darmawan2021a}, where the gate is based on beam splitter-type interactions rather than two-mode squeezing interactions. 
In this case, the STA does not work effectively, as shown recently~\cite{Kanao2024}.

\subsection{Cancellation term}
To cancel the additional AC-Zeeman shift in Eq.~(\ref{eq:stat_model}), we apply an additional DC-flux pulse to KPO1. 
This implementation corresponds to modifying Eq.~(\ref{eq:cancellation}) as follows:
\begin{equation}\label{eq:cancellation2}
\begin{split}    
&\theta_0-\delta_\textrm{m1}(t)-\theta_1(t) \\
&\qquad
\rightarrow
\theta_0-\delta_\textrm{m1}(t)-\theta_1(t)-\theta_\textrm{c}(t), 
\end{split}
\end{equation}
where $\theta_\textrm{c}(t)$ is the term to cancel the additional shift. 
We call it a cancellation term.

\subsection{Results}
\begin{figure}
    \centering
\includegraphics[width=0.45\textwidth]{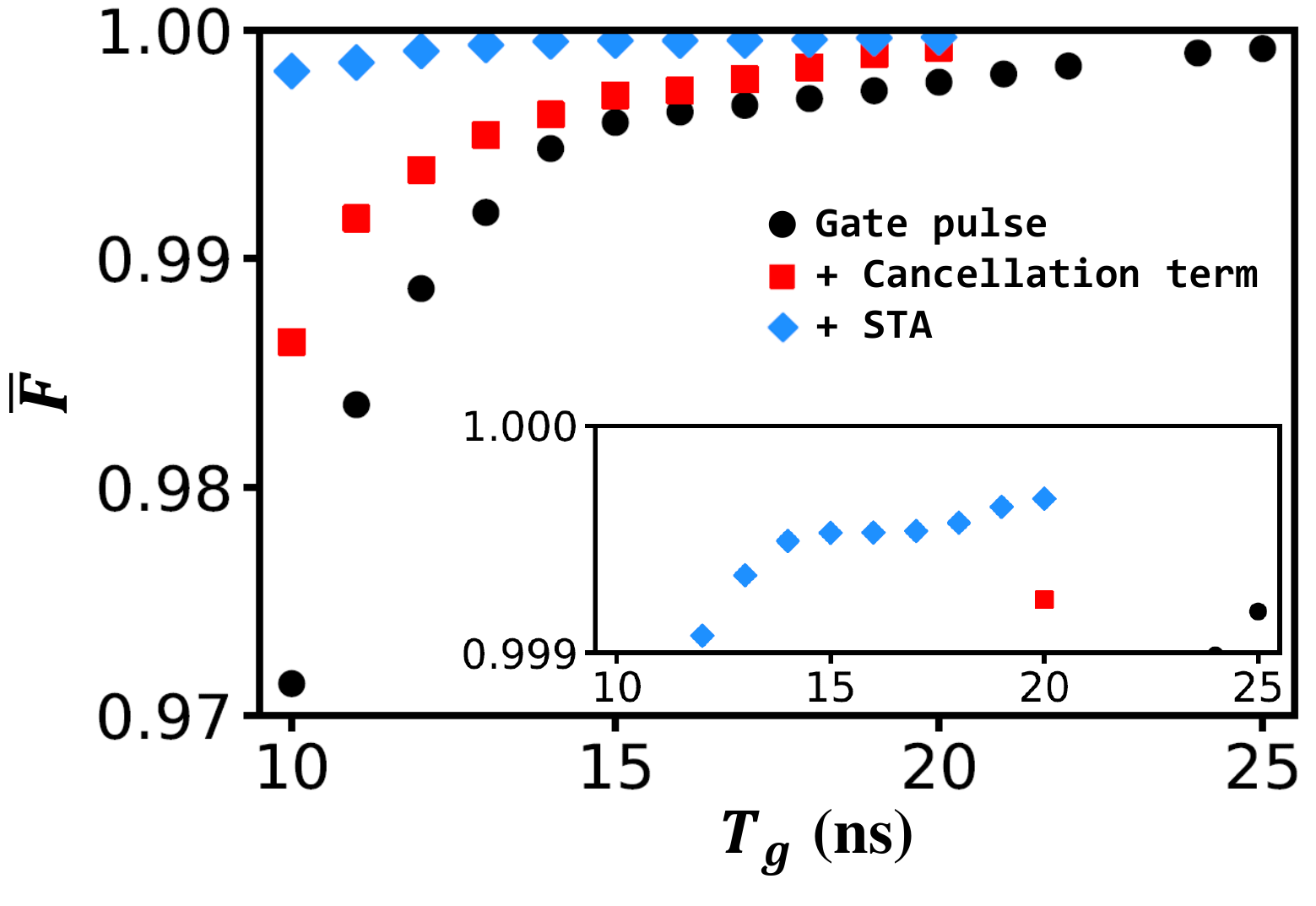}
\caption{\label{fig:ABC}
Average gate fidelity the CD gate with a cancellation term.
Black circles represent the results with the original gate pulse alone, red squares do the results with a cancellation term, and blue diamonds do the rusults with both the cancellation term and the counterdiabatic term for the STA. 
The inset shows a magnified view of $\bar{F}>$ 99.9\%.
} 
\end{figure}

\begin{figure*}
  \centering
\includegraphics[width=\textwidth]{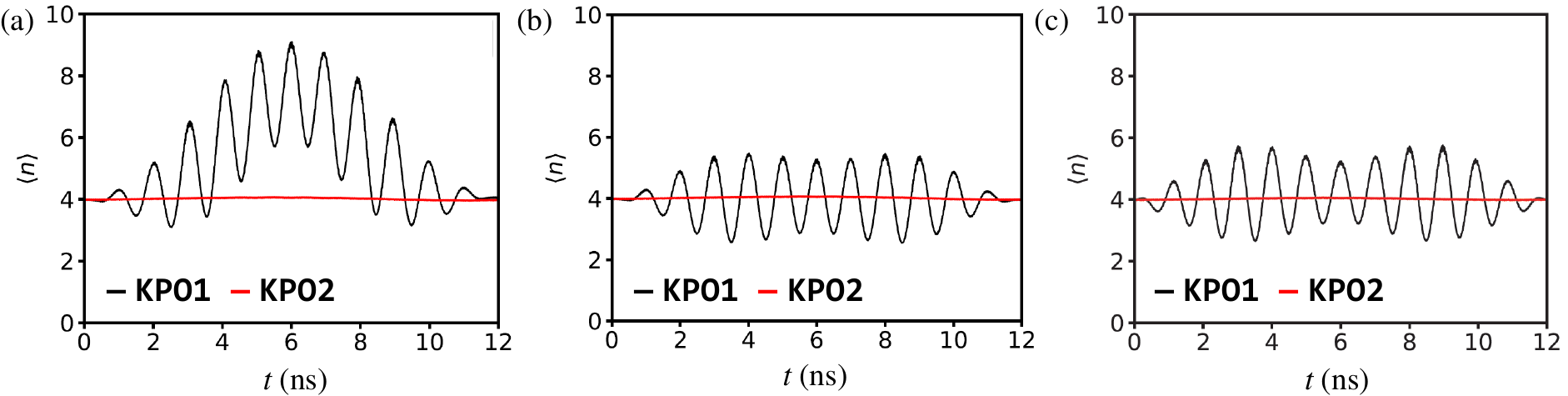}
\caption{\label{fig:num}
Average photon number of the KPOs during CD-gate operations.
(a) The original gate pulse alone.
(b) The original gate pulse with the cancellation term.
(c) The original gate pulse with the cancellation term and the counterdiabatic term for the STA.
} 
\end{figure*}  

\begin{figure*}
  \centering    
\includegraphics[width=\textwidth]{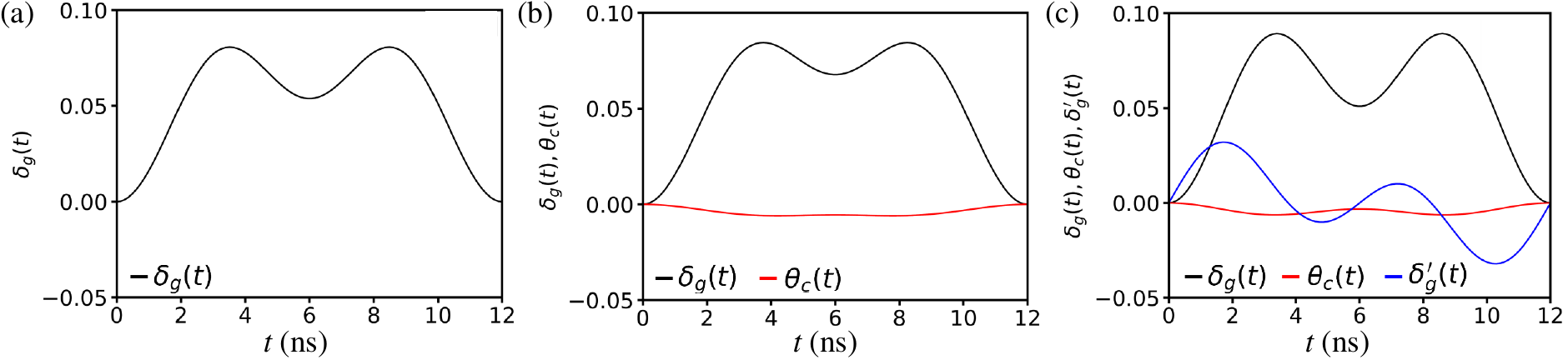}
\caption{\label{fig:pulse}
Optimized waveforms of the flux pulses for CD-gate operations.
(a) The original gate pulse alone.
(b) The original gate pulse (black) with the cancellation term (red).
(c) The original gate pulse (black) with the cancellation term (red) and the counterdiabatic term for the STA (blue).
} 
\end{figure*}

In a similar manner to the optimization in Sec.~IIC, 
we express the waveform of the cancellation term as
\begin{align}
\theta_\textrm{c}(t)
&=\sum_{n=1}^{N_\textrm{f}}\frac{C_n}{2}\left(1-\cos\frac{2n\pi t}{T_\textrm{g}}\right), 
\end{align}    
and optimize the parameters $A_n, B_n$, and $C_n$ of the pulse waveforms to maximize the average gate fidelity $\bar{F}$ in Eq.~(\ref{eq:Fbar}).
The results are shown in Fig.~\ref{fig:ABC}, from which it turns out that the shorter the gate time is, the more effective the cancellation term and the STA are for suppressing errors.
Focusing on average gate fidelities over 99.9\%, we find that with the original gate pulse alone, the gate time is 25~ns, but with the cancellation term, the gate time can be reduced to 20~ns, and further incorporation of the STA allows for its reduction to 12~ns. 
Thus, we have successfully shorten the gate time to less than a half of the original gate time.

The time evolutions of the average photon number are shown in Fig.~\ref{fig:num}. 
With the original gate pulse alone, the average photon number becomes unstable, increasing from 4 to 9. 
By adding the cancellation term, the average photon number becomes stable around 4.
This stabilization indicates that the increase in average photon number is caused by an additional AC-Zeeman shifts and suppressed by the cancellation term. 
For {$T_\textrm{g}=12$~ns}, the maximum value of the shift is estimated to be appropriately 50~MHz using $\delta_\textrm{g}(t)$ and $\delta^{\prime}_\textrm{g}(t)$ in Fig.~\ref{fig:pulse}(c) and Eqs.~(\ref{eq:pg}) and (\ref{eq:stat_model}).
Figure~\ref{fig:pulse} illustrates the pulse waveforms for the original gate pulse, a cancellation term, and a counterdiabatic term. 
For experimental feasibility we have limited the number of frequency components $N_\textrm{f}$ to only 2, leading to the simple waveforms in Fig.~\ref{fig:pulse}, as expected.

\begin{figure}
\includegraphics[width=0.45\textwidth]{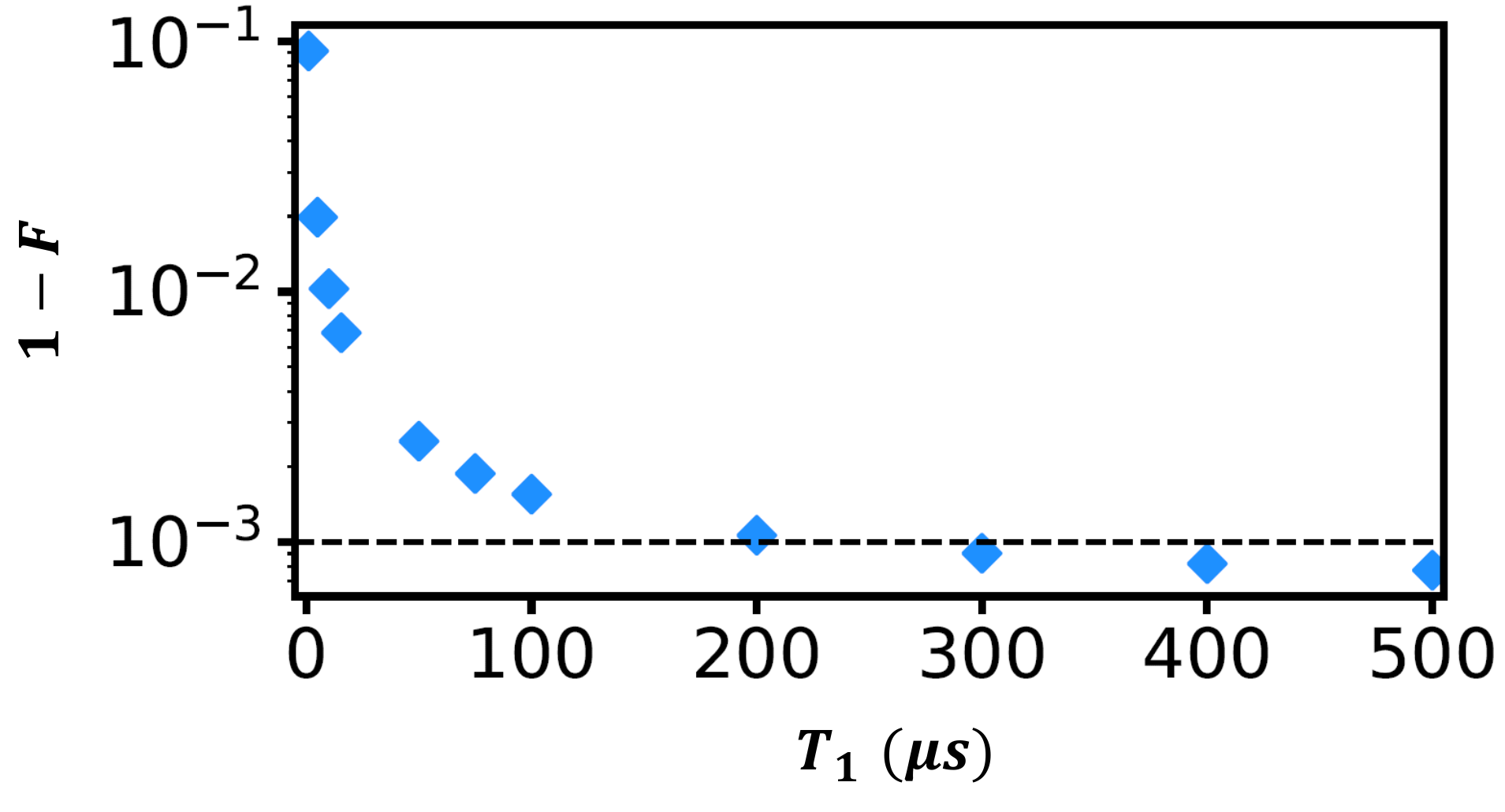}
\caption{\label{fig:inF}
Infidelity of CD gate with {$T_\textrm{g}=12$ ns} for the rotation angle $\pi/2$ corresponding to the single-photon rate {$T_1=1/\gamma$}.
The dashed line represents {$F=99.9\%$}.
}
\end{figure}
\section{Effect of single-photon loss}
Finally, we study the effect of single-photon loss in KPO qubits using the optimized flux pulses obtained in the previous section.
We evaluate the time evolution of the density operator $\rho$ by numerically solving the following master equation:
\begin{equation}\label{eq:master}
\partial_t\rho=-\im[H, \rho]+\frac{\gamma}{2}\sum_{j=1,2}(2a_j\rho{a}_j^\dagger-\rho{a_j^{\dagger}a_j}-{a_j^{\dagger}a_j}\rho), 
\end{equation}
where $\gamma$ is the single-photon loss rate ($1/T_1$) of KPO qubits.
Here we assume that the initial states of the two KPO qubits are cat states, 
$(\ket{\bar{0}}+\ket{\bar{1}})(\ket{\bar{0}}+\ket{\bar{1}})=\ket{\psi_\textrm{even}}+\ket{\psi_\textrm{odd}}$, where $\ket{\psi_\textrm{even}}=\ket{\bar{0}\bar{0}}+\ket{\bar{1}\bar{1}}$ and $\ket{\psi_\textrm{odd}}=\ket{\bar{0}\bar{1}}+\ket{\bar{0}\bar{1}}$,
and calculate the CD-gate infidelity for the rotation angle $\pi/2$ as a function of the decay time $T_1$, ranging from 1~$\mu$s to 500~$\mu$s. 
The fidelity is given by 
\begin{align}
F&=\braket{\psi_\textrm{id}|\rho(T_\textrm{g})|\psi_\textrm{id}}, \\
\ket{\psi_\textrm{id}}&\propto
\ket{\psi^{\prime}_\textrm{even}}+\e^{\im\tfrac{\pi}{2}}\ket{\psi^{\prime}_\textrm{odd}},
\end{align}
where $\ket{\psi^{\prime}_\textrm{even}}~(\ket{\psi^{\prime}_\textrm{odd}})$ is the final state obtained by time evolution starting from the initial state $\ket{\psi_\textrm{even}}~(\ket{\psi_\textrm{odd}})$ with {$\delta_\textrm{g}=\delta^{\prime}_\textrm{g}=\theta_\textrm{c}=0$}.
To reduce the computational cost for solving the master equation~(\ref{eq:master}), we perform an approximation by neglecting the higher-order terms than the 8th in the cosine matrix within Eq.~(\ref{eq:SC_model2}). 
We have confirmed that, in the case of no loss, the fidelities with and without the approximation are consistent by solving the corresponding Schr\"{o}dinger equation.
Figure~\ref{fig:inF} shows the infidelity for the gate time {$T_\textrm{g}=$12~ns}, indicating that a decay time of longer than 250~$\mu$s is required to keep the infidelity below 0.1$\%$.

\section{conclusion}\label{conc}
In this study, by deriving a static model from a superconducting-circuit model, we have found the undesired time-dependent AC-Zeeman shifts due to the strong AC-flux pulses for the CD gate, and have proposed the method to cancel the shifts with additional DC-flux pulses.
We have also numerically optimized the waveforms of a gate pulse, a counterdiabatic term for the STA, and the cancellation term.
As a result, the gate time of the CD gate has been reduced by approximately a half, from 25 ns to 12 ns, while keeping an average gate fidelity over 99.9\%.
Thus, we have shown that the CD gate for KPO qubits can be successfully accelerated by canceling the undesired AC-Zeeman shifts with additional DC-flux pulses. 
We have also studied the effect of single-photon loss. The result indicates the error probaility below 0.1$\%$ will require a decay time of longer than 250~$\mu$s.
We expect that our results will be useful for quantum computing with KPO qubits.

\section*{ackowledgements}
We thank T. Kanao for valuable discussion.
This paper is based on results obtained from Project No. JPNP16007,
commissioned by the New Energy and Industrial Technology
Development Organization (NEDO), Japan.
\appendix
\section{Derivation of the computaional basis states}\label{sec:appendA}
Here we derive appropriate computational basis states for the present model.
The amplitudes of coherent states of the KPO qubits are slightly deviated from those of isolated ones due to the interaction $V$ in Eq.~(\ref{eq:SC_model3}).  
To estimate stable ground states of the coupled two KPOs, we transform the bare operator $a_j$ into a new one $b_j$ as follows. 

We first diagonalize the linear part of the RWA model in the laboratory frame as
\begin{equation}
\label{eq:diag}  
  \begin{split}
H_\textrm{L}
&=\sum_{j=1,2}\omega_ja_j^{\dagger}a_j+g(a_1^{\dagger}a_2+a_1a_2^{\dagger}) \\
&=\begin{pmatrix}
  a_1^{\dagger} \\
  a_2^{\dagger}
  \end{pmatrix}^{\top}
  \begin{pmatrix}
  \omega_1 & g \\
  g & \omega_2 
  \end{pmatrix}
  \begin{pmatrix}
  a_1 \\
  a_2
  \end{pmatrix} \\
&\equiv 
 \omega_-b_1^{\dagger}b_1+\omega_+b_2^{\dagger}b_2,
  \end{split}
\end{equation}      
where 
\begin{equation}
\omega_\pm
=\frac{\omega_1+\omega_2\pm\sqrt{\Delta_{12}^2+4g^2}}{2}.
\end{equation} 
The unitary matrix diagonalizing the matrix in Eq.~(\ref{eq:diag}) is
\begin{align}\label{eq:Utilde}
\tilde{U}&=(\textbf{u}_1\quad\textbf{u}_2), \\
\textbf{u}_1^\top&=\mathcal{N}_1(-g\quad\omega_1-\omega_-), \nonumber \\
\textbf{u}_2^\top&=\mathcal{N}_2(-g\quad\omega_1-\omega_+), \nonumber
\end{align}
where $\mathcal{N}_1$ and $\mathcal{N}_2$ are normalsization factors.
Then, the annihilation operators, $a_1$ and $a_2$, are transformed into the new ones, $b_1$ and $b_2$, as
\begin{equation}\label{eq:b}
a_1 = \tilde{U}_{11}b_1+\tilde{U}_{12}b_2,\quad 
a_2 = \tilde{U}_{21}b_1+\tilde{U}_{22}b_2,
\end{equation}  
where $\tilde{U}_{ij}$ is the element of $\tilde{U}$ in Eq.~(\ref{eq:Utilde}).
Substituting Eq.~(\ref{eq:b}) into the RWA model [Eqs.~(\ref{eq:H1}-\ref{eq:g})]
and moving into the rotating frame with the unitary operator $\textrm{exp}[-\im(\omega_-b_1^{\dagger}b_1+\omega_+b_2^{\dagger}b_2)]$,
we obtain a $b$-mode RWA model during idle time:
\begin{align}\label{eq:RWA_b}
H_{\textrm{RWA}}^b
&=\sum_{j=1,2}\left[-\frac{K_j^b}{2}b_j^{\dagger 2}b_j^2+\frac{P_j^b}{2}(b_j^{\dagger 2}+b_j^2)\right] \nonumber \\
&\quad
-K_{12}^{b}b_1^{\dagger}b_2^{\dagger}b_1b_2,
\end{align}
where
\begin{align}
K_j^b&=K_1\tilde{U}_{1j}^4+K_2\tilde{U}_{2j}^4,~P_j^b=P_j\tilde{U}_{jj}^2, \\
K_{12}^b&=2(K_1\tilde{U}_{11}^2\tilde{U}_{12}^2+K_2\tilde{U}_{21}^2\tilde{U}_{22}^2).
\end{align}
Since $|\tilde{U}_{jj}|\gg|\tilde{U}_{ij}|~(i\neq~j)$, the stable coherent states for the $H_\textrm{RWA}^b$ in Eq.~(\ref{eq:RWA_b}) are appropriately given by $\ket{\pm\beta_j}$, where
\begin{align}
\beta_j=\sqrt{\frac{P_j^b}{K_j^b}}\simeq\frac{\alpha_j}{|\tilde{U}_{jj}|},~
\alpha_j=\sqrt{\frac{P_j}{K_j}}.
\end{align}

Since these two coherent states are not strictly orthogonal, orthogonalization and renormalization are necessary to use them as a computational basis. 
The orthogonal cat states for KPO$j$ are represented as
\begin{equation}
\ket{\mathcal{C}_j^{\pm}}=\frac{\ket{\beta_j}\pm\ket{-\beta_j}}
{\sqrt{2\pm2\braket{\beta_j|{-\beta_j}}}}.
\end{equation}
Using these, the computational basis states for a qubit can be written as
\begin{equation}
\ket{\bar{0}}=\frac{\ket{\mathcal{C}^+}+\ket{\mathcal{C}^-}}{\sqrt{2}},
\quad
\ket{\bar{1}}=\frac{\ket{\mathcal{C}^+}-\ket{\mathcal{C}^-}}{\sqrt{2}}.
\end{equation}
Thus, the computational basis states for two qubits are given by (four sign correspondence)
\small
\begin{widetext}
\begin{equation}\begin{split}\label{eq:basis}
&\begin{rcases}
    \ket{\bar{0}\bar{0}} \\
    \ket{\bar{0}\bar{1}} \\
    \ket{\bar{1}\bar{0}} \\
    \ket{\bar{1}\bar{1}}
\end{rcases}
=\frac{1}{2}
\left(
\ket{\mathcal{C}_1^+}\ket{\mathcal{C}_2^+}
\begin{matrix}
+ \\
- \\
+ \\
-
\end{matrix}
\ket{\mathcal{C}_1^+}\ket{\mathcal{C}_2^-}
\begin{matrix}
+ \\
+ \\
- \\
-
\end{matrix}
\ket{\mathcal{C}_1^-}\ket{\mathcal{C}_2^+}
\begin{matrix}
+ \\
- \\
- \\
+
\end{matrix}
\ket{\mathcal{C}_1^-}\ket{\mathcal{C}_2^-}     
\right) \\
&=\frac{1}{4}
\left(
\frac{\ket{\beta_1}\ket{\beta_2}+\ket{\beta_1}\ket{-\beta_2}
+\ket{-\beta_1}\ket{\beta_2}+\ket{-\beta_1}\ket{-\beta_2}}
{\sqrt{(1+\braket{\beta_1|{-\beta_1}})(1+\braket{\beta_2|{-\beta_2}})}} 
\begin{matrix}
+ \\
- \\
+ \\
-
\end{matrix}
\frac{\ket{\beta_1}\ket{\beta_2}-\ket{\beta_1}\ket{-\beta_2}
+\ket{-\beta_1}\ket{\beta_2}-\ket{-\beta_1}\ket{-\beta_2}}
{\sqrt{(1+\braket{\beta_1|{-\beta_1}})(1-\braket{\beta_2|{-\beta_2}})}}  \right.\\
&\quad\left.
\begin{matrix}
+ \\
+ \\
- \\
-
\end{matrix}
\frac{\ket{\beta_1}\ket{\beta_2}+\ket{\beta_1}\ket{-\beta_2}
-\ket{-\beta_1}\ket{\beta_2}-\ket{-\beta_1}\ket{-\beta_2}}
{\sqrt{(1-\braket{\beta_1|{-\beta_1}})(1+\braket{\beta_2|{-\beta_2}})}} 
\begin{matrix}
+ \\
- \\
- \\
+
\end{matrix}
\frac{\ket{\beta_1}\ket{\beta_2}-\ket{\beta_1}\ket{-\beta_2}
-\ket{-\beta_1}\ket{\beta_2}+\ket{-\beta_1}\ket{-\beta_2}}
{\sqrt{(1-\braket{\beta_1|{-\beta_1}})(1-\braket{\beta_2|{-\beta_2}})}} \right) \\
&=\frac{1}{4}
\left[
\left(
\mathcal{N}_{++}
\begin{matrix}
+ \\
- \\
- \\
+
\end{matrix}
\mathcal{N}_{+-} 
\begin{matrix}
+ \\
+ \\
- \\
-
\end{matrix}
\mathcal{N}_{-+} 
\begin{matrix}
+ \\
- \\
- \\
+
\end{matrix}
\mathcal{N}_{--}
\right)
\right.
\ket{\beta_1}\ket{\beta_2} 
+\left(
\mathcal{N}_{++}
\begin{matrix}
- \\
+ \\
- \\
+
\end{matrix}
\mathcal{N}_{+-} 
\begin{matrix}
+ \\
+ \\
- \\
-
\end{matrix}
\mathcal{N}_{-+} 
\begin{matrix}
- \\
+ \\
+ \\
-
\end{matrix}
\mathcal{N}_{--}
\right)
\ket{\beta_1}\ket{-\beta_2} \\
&\qquad
+\left(
\mathcal{N}_{++}
\begin{matrix}
+ \\
- \\
+ \\
-
\end{matrix}
\mathcal{N}_{+-} 
\begin{matrix}
+ \\
+ \\
- \\
-
\end{matrix}
\mathcal{N}_{-+} 
\begin{matrix}
- \\
+ \\
+ \\
-
\end{matrix}
\mathcal{N}_{--}
\right)
\ket{-\beta_1}\ket{\beta_2} 
\left.
+\left(
\mathcal{N}_{++}
\begin{matrix}
- \\
+ \\
- \\
+
\end{matrix}
\mathcal{N}_{+-} 
\begin{matrix}
- \\
- \\
+ \\
+
\end{matrix}
\mathcal{N}_{-+} 
\begin{matrix}
+ \\
- \\
- \\
+
\end{matrix}
\mathcal{N}_{--}
\right)
\ket{-\beta_1}\ket{-\beta_2}
\right],
\end{split}
\end{equation}
\end{widetext}
\normalsize
where 
\begin{equation}\label{eq:N}
\begin{split}
\mathcal{N}_{\pm\pm}=
\left[(1\pm\braket{\beta_1|{-\beta_1}})(1\pm\braket{\beta_2|{-\beta_2}})\right]^{-\frac{1}{2}}.
\end{split}
\end{equation}
Note that they are represented for the new modes $b_j$. 
Finally, to return to the bare modes $a_j$, we transform the basis states
${\ket{\pm\beta_1}\ket{\pm\beta_2}}$ in Eq.~(\ref{eq:basis}) to ${\ket{\alpha_{1}^{\pm\pm}}\ket{\alpha_2^{\pm\pm}}}$, where ${\alpha_{j}^{\pm\pm}=\pm\tilde{U}_{j1}\beta_1\pm\tilde{U}_{j2}\beta_2}$.
Using the resultant basis states, we calculate the average gate fidelity in Eq.~(\ref{eq:Fbar}) with Eq.~(\ref{eq:U}).

\section{Derivation of the static model}\label{sec:appendB}
We derive the static model [Eq.~(\ref{eq:stat_model})] from the RWA model by performing a high-frequency expansion with respect to $\Delta_{12}$.
Moving into the rotating frame with the unitary operator ${\bar{U}(t)=\e^{-{\im}H_\textrm{KPO}t}}$ as ${\ket{\psi(t)}=\bar{U}(t)\ket{\phi(t)}}$, the Schrödinger equation then becomes
\begin{equation}\label{eq:d_phi}
\partial_t\ket{\phi(t)}=-\im [O'_\textrm{t}\e^{-\im\Delta_{12}t}+O_\textrm{t}^{\prime\dagger}\e^{+\im\Delta_{12}t}]\ket{\phi(t)},
\end{equation}
where $O'_\textrm{t}=\bar{U}^{\dagger}(t)O_\textrm{t}\bar{U}(t)$.
Next, by integrating both sides with respect to time and repeatedly applying integration by parts, we obtain
\begin{align}\label{eq:phi_t}
\ket{\phi(t)}-\ket{\phi(0)} 
&\simeq\frac{O^{\prime}_\textrm{t}\e^{-\im\Delta_{12}t}-O_\textrm{t}^{\prime\dagger}\e^{\im\Delta_{12}t}}{\Delta_{12}}\ket{\phi(t)}\nonumber \\
&\quad
-\frac{O^{\prime}_\textrm{t}-O_\textrm{t}^{\prime\dagger}}{\Delta_{12}}\ket{\phi(0)}
+\mathcal{O}(\Delta_{12}^{-2}).
\end{align}
Substituting this into the right-hand side of Eq.~(\ref{eq:d_phi}), dropping the oscillating terms, and moving to the original rotating frame, we finally obtain the static model [Eq.~(\ref{eq:stat_model})]. 
Note that this derivation is similar to the high-frequency expansion in the Floquet theory, which can be applied when the Hamiltonian depends on time periodically~\cite{Mikami2016,Eckert2017,Shirai2013}.


\section*{references}


\end{document}